\newcommand{\beq}{\begin{equation}}
\newcommand{\eeq}{\end{equation}}
\newcommand{\tem}{}

\documentclass{aa}
                                                                                
\usepackage{graphicx}
                                                           
\begin{document}

\title{Fine structure, magnetic field and heating of sunspot penumbrae}

\date{Received (date) / Accepted (date)}

 \author{H.C.\ Spruit\inst{1}, \and G.B.\ Scharmer\inst{2} }
\institute{ Max Planck Institute for Astrophysics, Box 1317, 85741 Garching, Germany
\and
Institute for Solar Physics, Royal Swedish Academy of Sciences,
Alba Nova University Center, 10691 Stockholm, Sweden}                                                                               
\offprints{\\ H. Spruit \email{henk@mpa-garching.mpg.de}}
                        
\titlerunning{Penumbra structure, magnetic field and heating}
\titlerunning{The gappy penumbra}
\authorrunning{H.C.\ Spruit \& G.B.\ Scharmer}
                                     
\frenchspacing

\abstract{ 
We interpret penumbral filaments as due to convection in field-free, radially aligned gaps
just below the visible surface of the penumbra, intruding into a nearly potential field above. This
solves the classical discrepancy between the large heat flux and the low vertical velocities observed
in the penumbra. The presence of the gaps causes strong small-scale fluctuations in inclination,
azimuth angle and field strength, but without strong forces acting on the gas. The field is nearly
horizontal in a region around the cusp-shaped top of the gap, thereby providing an environment for
Evershed flows. We identify this region with the recently discovered dark penumbral cores. Its
darkness has the same cause as the dark lanes in umbral light-bridges, reproduced in numerical
simulations by Nordlund and Stein (2005). We predict that the large vertical and horizontal gradients
of the magnetic field inclination and azimuth in the potential field model will produce the net
circular polarization seen in observations. The model also explains the significant elevation of
bright filaments above their surroundings. It predicts that dark areas in the penumbra are of two
different kinds: dark filament cores containing the most inclined (horizontal) fields, and regions
between bright filaments, containing the least inclined field lines.
}

\maketitle

\section{Introduction}
\label{intro}
The  complex magnetic field structure and dynamics in sunspot penumbrae, in fact the very 
existence of penumbrae, present several outstanding puzzles in solar physics. This fine structure 
and its dynamics are evidently a consequence of unobservable sub--surface processes that we do not 
understand theoretically. 

One of the foremost theoretical problems associated with sunspot penumbrae (and also umbrae) is 
the heating problem: The bolometric brightness of the penumbra is some 75\% of the normal solar 
surface on average; even in the umbra it is still about 20\%. Carrying these heat fluxes requires 
large vertical velocities, of the order 1--2 km/s, which must also be of the right
correlation (upward hot, downward cool). The observations do not fit this requirement. The
problem is most serious in the umbra, where vertical velocity--intensity correlations are
quite low compared to what is needed to carry the observed heat flux (Beckers, 1977). The
velocities seen in the penumbra are larger, but mostly horizontal (Beckers and Schr\"oter
1969, Tritschler et al. 2004, Langhans et al. 2005a), with little upward motion in the bright 
components of the fine structure. There is thus a heat flux problem in the penumbra as well as 
in the umbra.

One obvious solution to the heat flux problem would be to assume that the penumbra is a very
shallow structure, such that the observed heat flux can be carried mostly by radiation from the
convection zone below. However, this would imply that the field in the penumbra is nearly
horizontal, which is not consistent with the observation that most of the magnetic flux of a
sunspot actually crosses the solar surface through the penumbra, not the umbra. The region below
the penumbra must be strongly magnetic, as in the quantitative `thick penumbra' models of Jahn and
Schmidt (1994). This rules out the shallow penumbra model of Schmidt et al. (1986). {\tem The 
observations also show that almost all the flux of the penumbra has the same sign as the umbra,  
and that only a small amount of flux of opposite polarity is in the form of moving magnetic features
outside at the the penumbra. This rules out the} recent model of Thomas et al. (2002), which 
interprets the structure of the penumbra as due to `turbulent pumping'.

In the case of the umbra, however, the heat flow problem has a well known solution: the spot's
apparently (at the surface) space--filling magnetic field actually contains a dense forest of
field--free gaps below the surface (Parker 1979a). The heat flux of the umbra is channeled through
these gaps by field--free convection. The contribution of the present paper is to take the logical
step of assuming that the penumbra is equally gappy below its observed surface. We show how,
besides solving the heat flow problem, this explains a number of other puzzling observations
brought into sharp focus by the recent high--resolution observations with the Swedish 1-m Solar
Telescope (Scharmer et al. 2002, Langhans et al. 2005a). {\tem Note that the interpretation of bright 
filaments as field--free regions just below the surface is in fact old, cf. Mamadazimov 
(1972)\footnote{On p134 of this paper, it says: `It seems to us that in all probability the bright 
interfilamentary elements are parts of the photosphere not covered by the dark filaments of the
penumbra'}}.

A major conceptual advantage of the model is that it provides a much more well defined framework 
for interpreting the observations than models referring more generically to some kind of 
magnetoconvection. Before we discuss the model, we briefly review some of the older and more 
recent observational evidence and the proposed interpretation (sects \ref{interp}, \ref{recinterp}).

In section \ref{model} we present a simple potential (current--free) field model for the penumbra, 
which takes into account field--free intrusions just below the visible surface. We show that
this leads to a magnetic field with strong fluctuations in inclination, azimuth angle and
strength above the surface without the need to invoke currents in the observed layers. This magnetic 
field structure is such that it allows locally horizontal or nearly horizontal magnetic fields, 
thereby providing a natural environment for Evershed flows (for which we do not claim to have an 
explanation, though we discuss the problem briefly in the discussion section.) We propose that the 
recently discovered dark cored penumbral 
filaments (Scharmer et al. 2002) represent the surface manifestation of this sub--surface 
convection and give observational and theoretical evidence to support this hypothesis. 

In the discussion section, we describe some properties of the model that appear consistent with
inferred properties of penumbral fine structure and magnetic fields, explain its relations with other
models, and make predictions that can be tested with current observational means.

\section{Existing interpretations of penumbral structure}
\label{interp}
Despite an overwhelming amount of data accumulated during the last decades,
a consistent description of penumbrae, based on observations, has failed
to emerge. It appears clear that a major problem in interpreting
these data is the small horizontal scales of these structures compared to the
spatial resolution of most observations, coupled with apparent rapid gradients with height 
in both the magnetic field and the Evershed flow. It is now
well established that the magnetic field shows strong fluctuations in inclination angle 
on small scales (e.g. Beckers and Schr\"oter 1969, Lites et al. 1990, Schmidt et al. 1992
and Title et al. 1993). Furthermore, it is generally agreed that Evershed flows are associated
with the more horizontal magnetic fields (e.g. Title et al. 1993) but conflicting evidence
exists on the correlation of penumbral intensity fluctuations with the inclination angle of
the magnetic field or the Evershed flow. For an extensive discussion of earlier observations,
we refer to the the review article of Solanki (2003).

In recent years, polarized spectra as well as measurements of broadband circular polarization, 
analyzed with inversion techniques, have given incontrovertible evidence for the existence of 
strong gradients or discontinuities along the line--of--sight in the magnetic field and 
the Evershed  flow, in particular along the vertical direction. This is most clearly
demonstrated by the lack of blue--red anti--symmetry of Stokes V profiles, resulting in a net
circular polarization when integrated over individual Zeeman sensitive spectral lines, as well as
broadband circular polarization when integrated over many such lines.  

Sanchez Almeida and Lites (1992) found that the Stokes profiles of their penumbral spectra 
required rapid changes in both the inclination of the magnetic field and in the flow speed. 
Their inversions indicated that the field is progressively 
more horizontal with depth, coupled with a flow that also increases with depth. These very large 
variations of the inclination angle with depth, on the order of 45--60$^\circ$, lead Solanki et al. 
(1993) and Solanki and Montavon (1993) to infer unacceptably large curvature forces that would 
be strong enough to destroy the sunspot, if globally present. Solanki and Montavon (1993) 
instead proposed a model wherein nearly horizontal flux tubes, imbedded in a more vertical 
magnetic field, gave rise to the observed net circular polarization. This model avoided the 
strong curvature (i.e. large changes in the inclination {\em along} the magnetic field) that
appeared to be indicated by the observations, a problem that was noted already by Sanchez Almeida
and Lites (1992), and emphasized by Solanki et al. (1993).

Sanchez Almeida (1998) pointed to discrepancies in various estimates of inclination changes and
claimed that these were consistent only with measurements of broad band circular polarization if
the typical scales of inclination changes are on the order 1--15~km. This conclusion was contested
by Martinez Pillet (2000) in favor of the imbedded flux tube model, further developed by him and by
Schlichenmaier and by Collados (2002), and inspired by simulations of Schlichenmaier et al.
(1998a,b). Subsequent investigators, including e.g M\"uller et al. (2002), Bellot Rubio et al. (2003,2004), 
Borrero et al. (2005), have used forward modelling or inversion techniques to confirm the ability of 
the embedded flux tube model to explain the observed penumbral Stokes profiles.{\tem In the 
inversions of Borrero et al. (2005)  the imbedded flux tube is represented in a very simplistic 
way, however. The perturbation of the background magnetic field, that is an unavoidable 
consequence of the imbedded flux tube (Solanki and Montavon 1993), is ignored. Moreover, 
only radiation from a single ray going through the center of the flux tube is calculated in the inversions, 
implying that the flux tube is actually modelled as a flux sheet with constant thicknes and without 
horizontal boundaries.  

It is also} clear that inversions applied to observed data do not have a unique interpretation.
For example, Westendorp Plaza et al. (2001a,b) found that the magnetic field strength {\em
increases} with height in the atmosphere, whereas Martinez Pillet (2000) emphasized that observed
profiles, obtained at 1" or lower resolution, are the results of a combination of profiles from
very different atmospheric conditions. He demonstrated that the results of the inversions of
Westendorp Plaza et al. (2001a,b), referred to above, can be obtained from a flux tube imbedded in
a more vertical magnetic field strength having a field strength {\em decreasing} with height.

Bellot Rubio et al. (2003, 2004) recorded Stokes spectra in the near infrared obtained from a 
symmetric
sunspot at 40$^\circ$ heliocentric distance. These authors applied one-- and two--component
inversions to the data and obtained good fits with two separate magnetic field components, one of
which was aligned with the Evershed flow, having inclination angles differing by more than
30$^\circ$ in the outer penumbra. These investigations provide indisputable evidence for the
existence of strong inhomogeneities, in the azimuthal direction as well as along the
line--of--sight, in penumbrae. However, they also demonstrate our inability to infer the detailed
nature of these inhomogeneities from observed Stokes spectra obtained at low spatial resolution.

 This difficulty is further highlighted by recent work of Borrero et al. (2004), who also
inverted their observational data in terms of a one--component model (or rather, a model with
components stratified along the line of sight) and a two--component model (in which the data are
interpreted as an average of two horizontally separated components). Both kinds of model give very
similar results. While this does confirm that a mixture of components is needed, it also makes
clear that very little can be deduced about the geometrical ordering of these components.

In their simulations, Schlichenmaier et al. (1998a,b) modelled a penumbral filament as a thin flux
tube, based on the siphon mechanism proposed by Meyer and Schmidt (1968). These simulations show
that a flux tube initially located at the magnetopause becomes buoyant as a result of radiative
heating by the underlying hotter quiet sun. The tube develops an upflow that bends horizontally and
continues outward from the center of the spot. Radiative cooling at the photosphere builds up a
pressure gradient that accelerates this flow outwards, explaining the Evershed flow. This is in
good overall agreement with inferred observed properties of flux tubes, described above.

Schlichenmaier et al. (1999) estimated cooling times from radiative transfer calculations and used
these to interpret bright penumbral grains as the result of hot horizontal flows in flux tubes,
cooling off radiatively in the penumbral photosphere. However, these flux tubes can only with the
greatest difficulty provide the heat needed to compensate the radiative losses of the entire
penumbra. Schlichenmaier and Solanki (2003) estimated, on the basis of an assumed upflow velocity
of 4 km/s, a flux tube diameter of 100 km and a temperature of the upflow of 12000 K, that such a
flow could provide the needed heat along a length of approximately 1200 km, after which the flux
tube must exit the penumbra or submerge within it and a new flux tube emerge.

{\tem This model thus implies a large amount of magnetic flux dipping down within the penumbra.} 
Whereas there is some evidence for downflows and return flux within the outer penumbra (Westendorp 
Plaza et al. 1997), this appears to be only a small fraction of the flux emerging through the penumbra 
(Solanki 2003). The highly resolved magnetograms analyzed by Langhans et al. (2005a) also show 
{\tem very few examples} of opposite polarities within penumbrae, of even very large sunspots. {\tem 
There is thus no evidence for significant return flux. Even the opposite polarities observed near the 
edge of the penumbra account for only a small fraction of the penumbral magnetic flux.} 

We conclude that although the simulations of Schlichenmaier et al. (1998a,b) provide a plausible
scenario for Evershed flows, it appears highly unlikely that such flows can provide the needed
uniform heating of the penumbra along its {\tem length} indicated by observations, and that
therefore this is not a likely mechanism for explaining the heating of the penumbra.

{\tem In the model proposed by Thomas et al. (2002), field lines emerging in the penumbra are kept 
submerged outside the spot by turbulent, compressible convection.  While such a mechanism 
might cause local variations in inclination near the edge of the penumbra, it can not explain the large 
variations in inclination seen throughout the penumbra, up to the transition to the umbra. The 
fluctuations in the field caused by opposite polarities around the spot decline rapidly with distance
from their source. In particular, they cannot possibly explain the large variations in inclination angle 
within the {\em height} over the first one or two hundred km above the penumbra photosphere required 
for consistency with observations. Moreover, this mechanism implies a significant amount of flux having 
opposite polarity to that of the penumbra in the photosphere surrounding sunspots, whereas very small
amounts of such flux is actually observed. The turbulent pumping mechanism proposed by Thomas
et al. (2002) is not consistent with these observations and can therefore not be responsible for the structure 
of the penumbral magnetic field. }

\section{Recent observations of fine structure of penumbral filaments}
\label{recinterp}
\begin{figure*}[tbh]
\includegraphics[width=\linewidth]{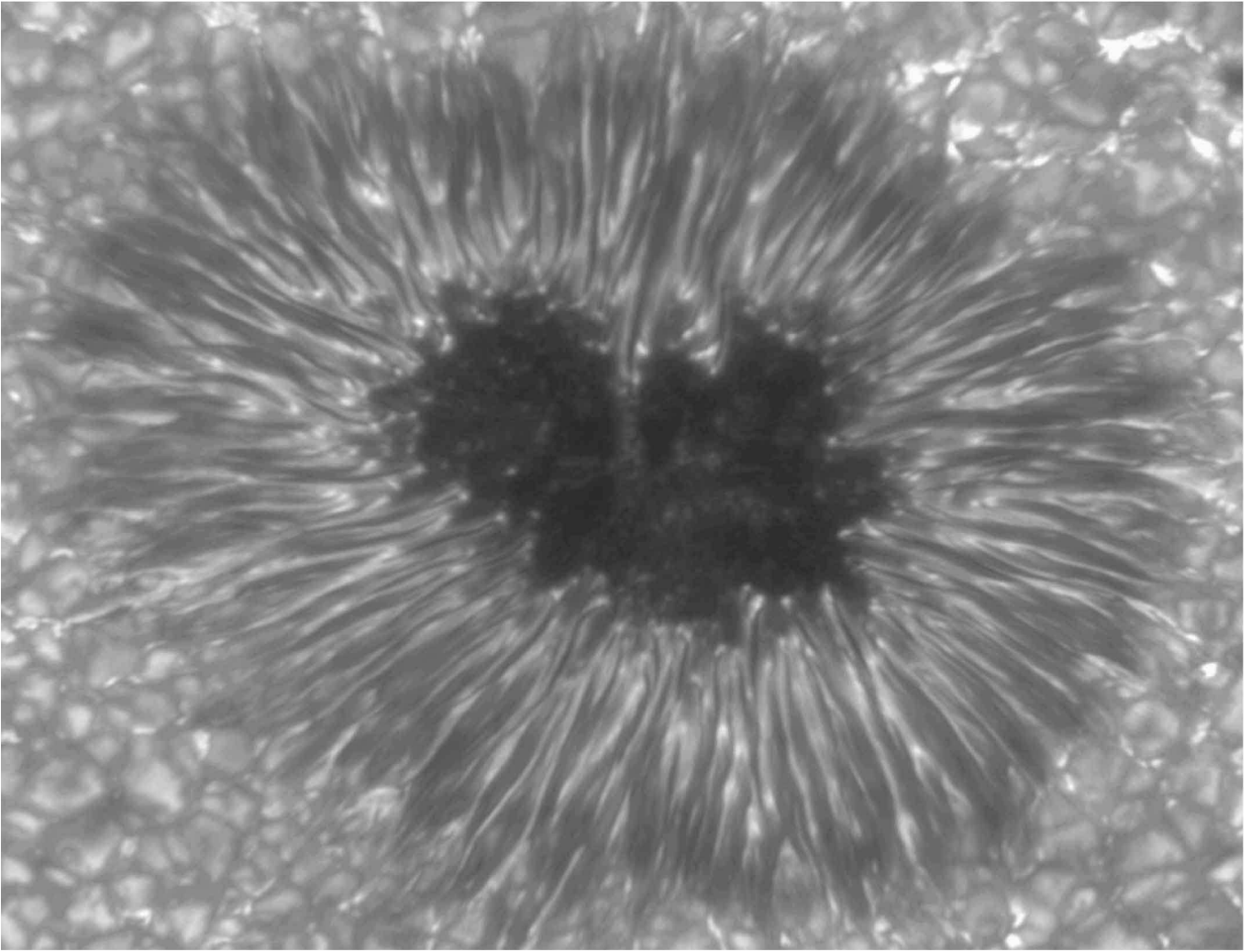}
     \caption{Sunspot observed with a circular polarizer in the red wing of the 630.25~nm
      Fe I line, showing dark cored penumbral filaments with exceptional clarity. More than 40 of
      these dark cored filaments originate in the umbra, most of them are associated with an umbral
      dot. (Image recorded by G\"oran Scharmer and Kai Langhans)
             }
     \label{cores1}
\end{figure*}

The most striking discovery made so far with the Swedish 1-m Solar Telescope (SST) 
has been that penumbral filaments consist of a dark core flanked by lateral brightenings 
(Scharmer et al. 2002). Apart from the potential diagnostics this offers,
the discovery indicates that we may be starting to resolve the fundamental scale of
penumbra filaments and that important clues to explaining them can be found from highly
resolved existing and future images and spectra. 

In Fig.\ \ref{cores1} we show an example of
a sunspot with dark cored filaments. This particular image was recorded with a circular
polarizer at the red wing of the 630.2~nm Fe I line, emphasizing the structure of the
filaments. We can see more than 40 dark cored filaments protruding into the umbra but there
are also several such filaments that originate in the middle penumbra. Several filaments can be 
followed over more than 3000 km and the coherence of these structures along their lengths
is striking. Other SST data (Rouppe van der Voort et al. 2003) show examples of filaments that
are more than 6000~km long. In particular, there are no systematic intensity gradients along the length 
of the filaments shown in Fig.\ \ref{cores1}. This would be expected if these were individual flux 
tubes consisting of a hot upflow and a horizontal outflow, proposed by Schlichenmaier and Solanki 
(2003) to be responsible for the heating of the penumbra. 

Associating the dark cored filaments with individual flux tubes imbedded in a more vertical
magnetic field, as suggested by Solanki and Montavon (1993) is possible, but would imply that such
flux tubes must follow very closely the $\tau=1$ surface to be visible over lengths of several 1000
km, and the same must be true for the Evershed flow and the magnetic field within the flux tube.
While we do not completely reject the possibility that these structures are individual flux tubes
containing siphon flows, we believe that the large radial extension of the dark cored filaments
makes this unlikely and definitely question the suggestion by Schlichenmaier and Solanki (2003)
that such flows provide the heat flux to the penumbra. As discussed by Schlichenmaier and Solanki
(2003) we can also rule out interchange of flux tubes, proposed by Jahn and Schmidt (1994), as a
viable mechanism to heat the penumbra simply on the basis of the long life times of these dark
cored filaments, that preserve their identity during more than one hour (Langhans et al 2005a).

\begin{figure*}[tbh]
\includegraphics[width=\linewidth]{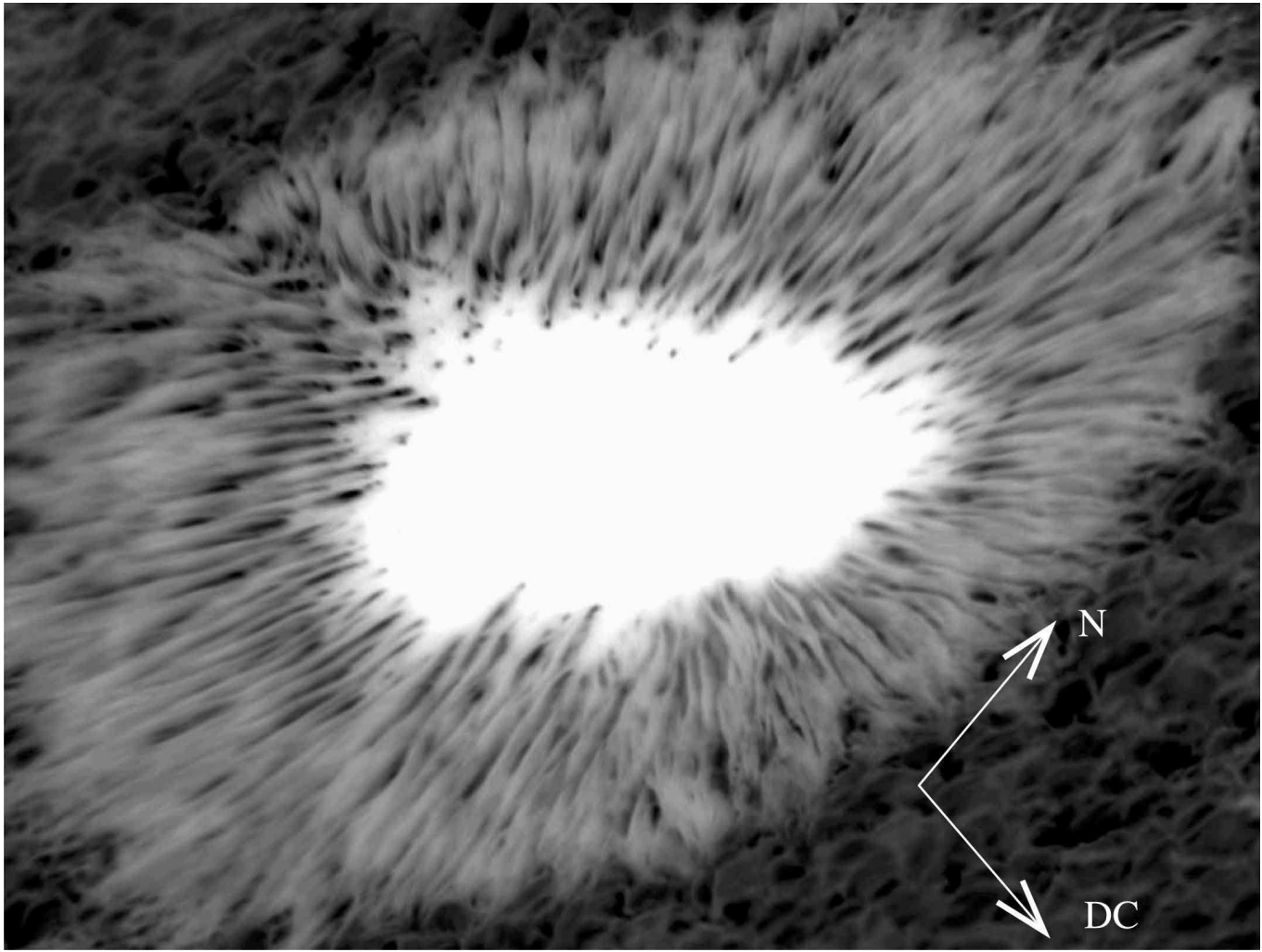}
     \caption{Evidence for 3D effects in sunspot penumbrae observed at a heliocentric distance of
     61$^\circ$ The image is shown negative to enhance visibility of dark penumbral cores, here shown
     brighter than the background penumbra intensity. The disk center (DC) and solar north (N) 
     directions are indicated. (Courtesy of Mats L\"ofdahl)
             }
     \label{cores2}
\end{figure*}

Fig.\ \ref{cores2} shows a continuum image of a fairly regular spot recorded at a heliocentric 
angle of 61$^\circ$ The disk center (DC) and solar North (N) directions are indicated. This image 
gives several indications suggesting that the $\tau=1$ surface in the penumbra is not a flat but a 
corrugated surface. Dark cores are not seen in most of the penumbra. On the limb side, dark 
cores are clearly seen in several filaments. {\tem On the center side, the dark cores are wider 
and of lower contrast than on the limb side.}

At azimuths away from the direction to the limb, dark cores shift away from the center of bright 
filaments towards the limb direction. At azimuths around 90 degrees, dark cores cannot easily be 
seen but hints of them are seen as dark streaks (shown as bright on this negative print)
separating two filaments. We also note that filaments in a narrow range of azimuth angles around 
the disk center and limb sides show markedly sharper structures than the more fuzzy filaments seen
away from these directions. This is not an artifact of differential seeing but can be seen
consistently in the more than one hour long movie produced from this data set. Finally, we note
that filaments near the limb side of the penumbra are associated with dot--like brightenings, as
are seen in sunspot images recorded near sun center, but that such brightenings are absent in
filaments near the sun center direction. This absence of brightenings on the Sun center side was
noted also by Tritschler et al. (2004) in a sunspot only 23$^\circ$ off disk center.

This and other images of sunspots far away from disk center recorded with the SST show similar
evidence of a complex, $\tau=1$ surface where the appearance of the penumbra filaments depends
strongly on the azimuth of the viewing angle. Interpreting such images is a delicate problem that
is open to personal bias. Here, we give only a partial interpretation by suggesting that we see
elevated bright filaments and that the dark cores outline the center and top of these filaments.
The suggestion that bright filaments are elevated has been made recently by Schmidt and Fritz
(2004), based on the center--to--limb and azimuthal variation of the penumbra brightness for about
80 sunspots. These authors modeled the penumbra as consisting of bright filaments in the shape of
elevated bright "boxes" on a dark background. In this simple model that ignores radiative transfer
effects, the dark background between the filaments is partly or completely obscured by the bright
filaments, leading to modulation of the limb darkening and azimuthal intensity variations. These
authors also note that for heliocentric angles larger than 60$^\circ$, the intensity of the
penumbra at the limb side is higher than at the center side. This could be explained as an opacity
effect but also by the absence of penumbral brightenings on the disk center side, seen in Fig.\
\ref{cores2}. {\tem The interpretation of dark, but partly optically thin, cores lying `on top' of
flux tubes was confirmed by S\"utterlin et al. (2004) with data} from the Dutch Open Telescope
(DOT). These authors also noted the prominence of point--like penumbral grains on the limb side,
but not center side, of the sunspot observed at a heliocentric distance of only 27$^\circ$. This,
and the results of Tritschler et al. (2004) would indicate that the `head' of a penumbral filament
extending into the umbra is inclined by at least some 45$^\circ$ with respect to the horizontal
plane.

Elevation above the background has also been inferred for umbral dots by Lites et al (1991),
who noted the lack of foreshortening of umbral dots observed near the limb. This is consistent with 
the close connection between umbral dots and penumbral filaments
 
\section {The connection of penumbral filaments with umbral dots and light bridges}
\subsection{Connection with umbral dots}

\label{connectu}
The observed continuous evolution of many penumbral filaments into umbral dots (Muller 1973,
Scharmer et al. 2002), is illustrated in Fig.\ \ref{cores1}. This shows more than 40 penumbral
filaments protruding into the umbra and in most cases connecting to a peripheral umbral dot.
This connection between umbral dots and penumbral filaments suggests common underlying 
physics.

For the umbra, the most reasonable solution to the heat flux problem is that the heat is
supplied in the form of radiation, from a source just below the observed umbral
surface. A key indication for this are the {\em umbral dots}. In this radiative heating
interpretation, a dense pattern of field-free gaps exists just below the umbral surface
which communicate with the surrounding convection zone. These gaps close just below
$\tau=1$, and the heating around their tips is seen as umbral dots. Collectively the dots
make up the net heat flux through the umbra.

The existence of these gaps was inferred theoretically (Parker 1979a) from the instability
of a vertical magnetic flux bundle to splitting (Meyer et al. 1977), which sets in just
below the observed surface. Based on this idea Spruit (1981) has proposed a model for a
sunspot in which a bundle of individual thin flux tubes is held together near the base of
the convection zone (cf. Spruit and Roberts 1983). The structure of the spot as seen at the
surface in this model reflects the balance between the magnetic buoyancy of the tubes (which
keeps them straight and vertical), and their mutual magnetic repulsion at the surface. When
the tubes are in temperature equilibrium with their surroundings, their field strength and
filling factor can be computed as a function of depth. These results also {\tem confirm} that gaps
must exist (filling factor less than 1) just below the umbral photosphere.

The division of a vertical bundle of field lines into slender strands separated by field
free convecting plasma is also assisted by the `convective expulsion' process (Zel'dovich
1956, Parker 1963, Weiss, 1966). A convecting magnetized plasma tends to separate into field
free overturning volumes with the magnetic field concentrated into isolated strands. The
strength of these `flux tubes' is just sufficient to resist further tangling and stretching
by the convective flow.  Once the field has separated into strands, a small amount of  fluid
motion in their environment is sufficient to keep them concentrated against the action of
magnetic diffusion.

The observed connection between umbral dots and penumbral filaments suggest that, just like an
umbral dot, a penumbral filament is in fact the surface manifestation of a field--free gap
below the observed surface, communicating directly with the surrounding convection zone. 

\subsection {The connection with light bridges}
Another discovery made with the Swedish 1m telescope was fine structure in umbral light bridges.
This fine structure consists both of short, narrow dark lanes oriented roughly perpendicular to the
axis of the light bridges and a long narrow dark lane running parallel to this axis (Lites et al.
2004). When viewed near the limb, the fine structure of such light bridges appear asymmetric with
respect to the limb and sun center directions, giving the impression of a raised structure against
a dark background. Assuming this and by taking advantage of the asymmetry of this structure between
the sun center and limb side directions of, Lites et al (2004) could estimate the apparent height
of these structures to be about 300 km. This is of the same order of magnitude as the Wilson
depression of sunspots calculated from models (Solanki et al. 1993). Such raised structures in the
umbra are expected for structures that have reduced magnetic field strength and thereby increased
gas pressure. However, we cannot infer increased gas pressure directly from the fact that a bright
structure appears elevated since this may be a consequence simply of the high temperature
sensitivity of the H$^-$ opacity.

Recently, Nordlund and Stein (2005) have made  3--D radiative magnetohydrodynamic
simulations that explain the dark lane running along the main axis of light bridges.  These
simulations started with two strong magnetic fields of the same polarity on either side of a
realistic field--free, convecting photospheric model. After allowing the model to relax from
this initial condition and calculating the emerging radiation, a narrow dark lane appears along
the symmetry axis between the two magnetic fields.  Viewing at different angles of incidence
shows this dark lane as a structure located somewhat above the center of the light bridge. The
simulations showed it to be unrelated to the convective flow pattern within or near the lane and
therefore not a signature of flows. The dark lane is actually a direct consequence of the higher
gas pressure in the region between the two magnetic fields which, together with the temperature
decrease with hight, shifts the $\tau=1$ surface upwards to cooler layers. The dark lane forms in
the field--free part of the photosphere below the cusp of the magnetic field. It simply outlines a
region of enhanced gas pressure compared with the surroundings.
 
Though light bridges are field--free at the level of the continuum photosphere
where the gas pressure can keep the strong fields separate, the rapid decrease of pressure
with height implies that these gaps close within {\tem some} hundred km above the photosphere, 
forming a cusp (Fig.\ \ref{gaps}, middle panel). Since this is only slightly above the surface, magnetic 
spectral lines formed 100 km or more above the photosphere should not show 
indications of a field--free region near the dark lane.  

The dark cores of bright penumbral filaments look similar to the dark lanes in light bridges. 
We take this as our second clue suggesting that they are likewise closely associated with narrow 
field-free zones between regions of strong magnetic field. 

\section{Penumbral energy balance: gaps in an inclined field} 
\begin{figure}
 \centering  \includegraphics[width=\hsize]{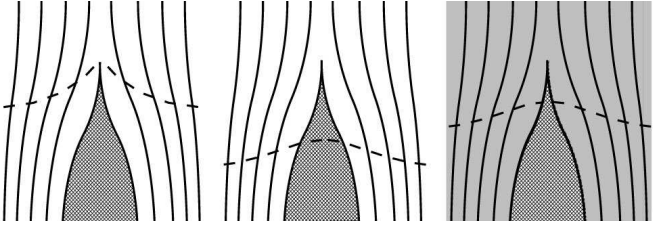}
\caption{\small Gaps (hatched) in a magnetic field near the solar surface (vertical
cross--sections). Dashed lines indicate the continuum $\tau=1$ level.  The two neighboring
flux bundles spread out horizontally above the surface, forming a cusp at some height above
$\tau=1$. Left: cusp is located below $\tau=1$, corresponding to an umbral dot. The surface
around the gap is brightened by the radiative heat flux. The observed field strength is
reduced due to the displacement of field lines by the gap. Middle: A wide gap that would be
seen as a field--free `canal' or umbral light bridge. Right: the case of a penumbral filament
in the proposed model is like a light bridge, but with an additional horizontal field
component  (strength indicated by shading) along the filament.
\label{gaps} }
\end{figure}

Based mainly on theoretical arguments
and primarily the absence of alternative scenarios that explain penumbral heating in accordance
with existing data in a satisfactory way, we propose that the penumbra is heated by field--free
convection occurring just below the visible surface. We interpret the apparent elevation of
penumbral filaments as evidence for enhanced gas pressure, caused by the reduced magnetic field
below the centers of such filaments (cf. the sketch in Fig.\ \ref{gaps}). 

Just like the light bridges computed by Nordlund and Stein discussed above, we interpret
the dark penumbral cores as signatures of the enhanced gas pressure  in the gap compared with
the magnetic field surrounding it. The symmetric lateral brightenings seen in penumbrae observed
near disk center constitute the bright and inclined sides of the elevated filaments. Their
brightness can be interpreted as due to a `bright wall effect' in exactly the same way as in
isolated photospheric flux tubes (Spruit 1976, 1977), recently reproduced in realistic 3--D
radiative magnetohydrodynamic simulations (Carlsson et al. 2004, Keller et al. 2004, Steiner 2005).

The optical depth in the magnetic field bounding the gap is reduced, so that one looks into the
field--free walls of the gap, down to some few hundred km below the nominal photospheric level. The
temperature in the convection zone at that depth is quite high, but as in the case of photospheric
flux tubes, radiative cooling of the walls reduces the temperature to something close to normal
photospheric temperatures. This explains the observed temperatures of the bright filaments in a way
which can be checked quantitatively by numerical simulations like those that have already been done
for photospheric flux tubes and light bridges.

The difference between the top and inclined sides of the filaments is dependent on the {\em
vertical} component of the magnetic field. When this vertical component is weak, as is the case in
the outer penumbra, the distinct difference in pressure balance, as well as in optical depth to the
gap, between the top and the sides of the filaments is less pronounced. This explains why the dark
penumbral cores are seen primarily in the inner and mid penumbra, where the {\em 
background} magnetic field is more vertical.

Through the presence of the magnetic field convection below the surface manifests itself, perhaps
somewhat counter--intuitively, as a combination of a dark and a bright structure, instead of just a
bright structure.

In our model, the main difference between light bridges and penumbral filaments is the presence of
a strong horizontal magnetic field component in the penumbra. We note in this context that in the
models of embedded flux tubes of Martinez Pillet (2000),  the flux tubes were assumed to have a
higher pressure and reduced field strength. This gave reduced continuum intensities irrespective of
whether the flux tubes were warmer or colder than the surroundings.

Although this interpretation of SST data is subject both to the possibility of personal bias and
the strong temperature sensitivity of the H$^-$ opacity, we believe that it is a highly plausible
scenario which is also strongly supported by the numerical simulations of Nordlund and Stein
(2005). In particular, our model allows uniform heating of filaments along their lengths by
convection below the visible surface, thus eliminating the need to provide such heating by
horizontal Evershed flows.

The scenario proposed here implies a penumbral magnetic field arranged in the form of radial
sheets below the surface, separated by field--free gaps. Above the surface, such a magnetic field 
must expand and `push over' the centers of the filaments, so that the field line inclination 
(angle with respect to the vertical) is increased. For the same reason, the field strength must
be somewhat lower than in the surroundings. 

A striking property of penumbral filaments is their inward motion, known to represent a pattern
motion rather than a fluid displacement (Muller 1973). In terms of our model, this means that the gaps
open up progressively towards the umbra. The exact reason for this opening process is a separate, 
though very interesting, question which we do not attempt to answer in this paper.

\section{Above the surface the field must be nearly current--free} 

\subsection{Forces in embedded tube models}
\label{forces}
Due to the very rapidly declining pressure in the atmosphere above the $\tau=1$ surface,
there are strong limits on the forces that can be present in the magnetic field configuration above
the photosphere (recall that the scale height of the atmosphere is only somewhat larger than the
spatial resolution of the best images taken with the Swedish 1--m Solar Telescope).

Observations such as those mentioned in section \ref{interp} require strong changes in direction
and/or strength of the field on small scales. Arbitrary variations in a magnetic field (i.e.
restricted only by $\mathrm{div}{\bf B}=0$) on a length scale $L$ are associated with currents of
the order $c B/(4\pi L)$, and volume forces of the order $B^2/(8\pi L)$. The gas pressure in the
layers observed in a {\tem penumbral} magnetogram is too small to balance these forces.

This problem can be alleviated by restricting the kind of variations in the fields. Since the
forces are due to currents perpendicular to field lines, they can be minimized by choosing fields
that are approximately {\em force free} (currents being parallel to the field lines). This
restriction, however, conflicts with the most popular models proposed for the inhomogeneities.

In the models of Solanki and Montavon (1993), Schlichenmaier et al. (1998a), Bellot Rubio et al.
(2005) flux tubes are embedded into a background field of a different direction. Such a
configuration can not be force free. This is easiest to see if the background field is
perpendicular to the tubes, but the argument carries over to general inclination. To accommodate the
tube (horizontal, say), the background field (vertical) must be `pushed aside', opening a gap. In
order for magnetic forces to be absent, the field strength must be continuous at the boundary
between the tube and the background field, and the field strength must also be constant across the
tube. At the top and bottom of the tube, this is not possible because the disturbed background
field vanishes there. This is more than just a mathematical inconvenience. The imbalance of forces
at top and bottom will cause the tube to flatten horizontally and expand vertically. This releases
the energy associated with the perturbation induced in the background field. The flattening happens
on an Alfv\'en crossing time: not more than a few tens of {\em seconds}, for the inferred densities
and length scales.

If the tube is allowed to flatten to completion, the process will eventually slow down and a
configuration of magnetic sheets of alternating direction is reached. Such a state can be
constructed in principle as a force free configuration. {\tem Such a sheet--like configuration }
has been proposed by Martens et al. (1996). This configuration, however, as well as any other 
force--free but not current--free configuration, generates a new, equally debilitating problem.

\subsection{force--free fields} 

A force--free configuration has {\em internal torques}. In any such
field these torques must be taken up by a surface to which the field lines are connected.  In
an isolated spot, almost all penumbral field lines return to the surface at a large distance from
the spot. The number of field lines dipping down into the photosphere at the edge of the penumbra
is small compared with the total magnetic flux crossing the surface through the penumbra. At
{\tem the distance where the penumbral field lines return to the surface}, the field strength is too low to 
take up the torques associated with any significant variations in direction {\tem due to 
field aligned currents in the penumbra}.

This argument can be made quantitative as follows. {\tem Let ${\bf B}_{\rm p}$ be} the untwisted
field (the field that would be there in the absence of currents), and ${\bf B}_{\rm t}$ the
component perpendicular to ${\bf B}_{\rm p}$ associated with the force--free currents {in a flux
bundle of width $l$}, {\tem the torque  on the bundle is constant along its length and 
proportional to $l B_{\rm p}B_{\rm t}/4\pi$.  Since $B_{\rm p}$ decreases strongly with distance 
from the center of the spot (like $1/r^2$, so that $l\sim 1/r$ ), the constancy of the torque 
implies that the ratio $B_{\rm t}/B_{\rm p}$ increases (as $r$) with distance from the spot.  This is 
a manifestation
of the well--known fact (e.g.\ Parker quoted above) that the twist in a force--free field accumulates
at the lowest field strengths encountered along a field line. Thus if something in the penumbra
twists field lines around each other, creating force--free currents, the twist thus induced propagates
away (at the Alfv\'en speed) to the place where the field is weak, i.e. into the corona. This makes 
it hard to maintain much twist in the penumbra itself. }

The differences in inclination seen at the edge of the penumbra are already of the order of one
radian. If these differences were due to force--free currents, they would have to be even larger
outside the penumbra. At a distance of only 2 spot radii, the differences in inclination
would already have to be of the order 2 radians. There is little room in the observations for such
large variations in inclination. The structure of the super--penumbra as seen in $H_\alpha$ for
example does not show show evidence of much variation in inclination at all, let alone differences
of the order 90$^\circ$. This problem also becomes apparent immediately as soon as one attempts to
extend a force--free field solution to any significant distance from the spot.

The conclusion is that field--aligned currents can not be used to explain the observed
variations of field inclinations. These variations must thus be due to currents perpendicular to
field lines. Since this is not possible in the low density regions where the spectral lines are
formed, the currents must instead be located deeper down. This is in fact the natural solution to
the puzzle of mixed field inclinations, as is shown in the following. 

\subsection{Varying inclinations in a potential field}

Variations in inclination on small scales as seen in the penumbra are also possible in
potential (= untwisted, current--free) fields. By the nature of scalar potentials,
irregularities decrease away from their boundaries. If the length scale of the
conditions imposed at the boundaries is $L$, the amplitude of the irregularities decreases
as $\exp(-z/L)$ with distance $z$ from the boundary. The observed length scale of the
irregularities in field inclination in the {\tem penumbra} is quite small, and their vertical length
scale must be equally small if the field is potential. There is, however, very little in the
observations that would contradict this. On the contrary, the interpretation of line polarization 
in spatially unresolved observations discussed above {\em requires} strong gradients in
inclination over the height of formation of the line as discussed in Sect. 2. This {\tem length} is 
of the same order or smaller than the horizontal length scales in the penumbra.

We turn this line of argument around: the observation of
strong variations in field inclination then must mean that the observed level in the
atmosphere is in fact quite close to the boundary where these irregularities are imposed.
This, of course, fits directly with the strong inhomogeneity just below the 
observed surface which we inferred from the heat flux problem (sect. \ref{intro}).

\section{A potential field model}
\label{model}
The above qualitative description can be made more precise with an exact potential field model. 
{\tem Such a field is current--free, hence has the form ${\bf B}=\nabla\phi$, where the scalar potential
$\phi$ satisfies the Laplace equation $\nabla^2\phi=0$. }Let
$z$ be the vertical coordinate, $y$ the radial horizontal component (perpendicular to the plane in
Fig. \ref{gaps}), and $x$ the tangential component (azimuthal in a frame centered on the spot
umbra). The model is two-dimensional, \beq  \partial_y =0 ,\eeq i.e. variations along the length
of the penumbral filament are ignored. The boundary conditions defining the field are imposed at
infinity and at the surface of the gap. At large distance from the gap the field is uniform. At the
surface of the gap the normal component of the field vanishes. We write the field as \beq {\bf
B}={\bf B}_\perp+{B_y}{\rm e}_y, \eeq where ${\bf B}_\perp$ is the field  in the $(x,z)$--plane,
perpendicular to the filament, and $B_y$ the parallel component. Consider first the perpendicular
components. A simple simple field satisfying the boundary conditions is

\beq 
{\bf B_\perp}={\bf B}_0 +{\bf B}_{\rm g},
\eeq
where ${\bf B}_0$ is a uniform vertical `background field' :
\beq B_{z0} {\bf\hat z},\eeq
while the field ${\bf B}_{\rm g}$, the disturbance due to the presence of the gap,
\beq B_{{\rm g}x}=\partial_x\phi; \qquad B_{{\rm g}z}=\partial_z\phi ,\eeq
has the (2--D) potential
\beq
\phi= \sin(k x) \exp(-k z).\eeq 
This is a simple periodic potential in the $x$--direction.{\tem  Note that by virtue of the properties
of scalar potentials,} the scale of its exponential decrease with height $z$ is tied to by the azimuthal 
separation $L=2\pi/k$ between the filaments. 

Like the background field ${\bf B}_0$, the parallel component is taken to be uniform. At
large distance from the gap, the field thus has a constant  inclination  $\alpha_0$ in the
$(y,z)$--plane:
\beq B_y/B_z\vert_\infty=\tan\alpha_0.\eeq

\begin{figure}[htbp]
 \centering  \includegraphics[width=1\hsize]{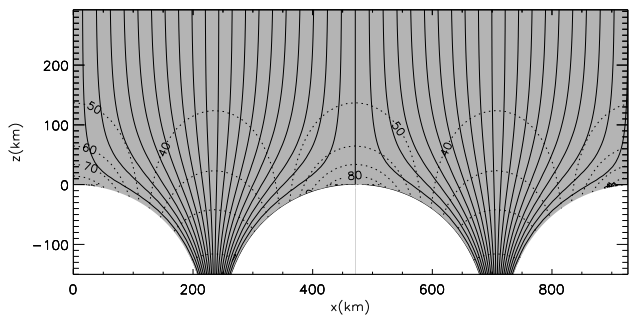}
 \caption{\small {\tem Idealized} field configuration of a penumbral filament. Solid lines show 
 the field lines projected onto a vertical ($x-z$) plane perpendicular to the filament. The 
 field-free region is shown in white. The ($y$-)component of the field (parallel to the filament,
perpendicular to the plane) is indicated by shading. It is uniform outside the gap, such
that far from the gap the inclination of the field (angle with respect to the vertical in
the $y-z$ plane) is 45$^\circ$. Dotted lines show contours of constant field line
inclination in the $(y,z)$-plane. The higher inclinations above the gap are due to the lower
field strength there (cf. Fig.\ \ref{B-inc}). [This model does not contain the cusp region shown 
in Fig.\ \ref{gaps}; it will appear, however, in models that take the pressure balance condition 
into account more realistically, such as those of Nordlund and Stein (2005)].}
\label{potfield}
\end{figure}

The simplicity of this field is made possible by assuming the freedom of specifying a
suitable shape for the surface of the gap. Its field lines are shown in Fig.\ \ref{potfield}. 

The model is associated with a discontinuity in the form of a current sheet at the boundary 
with the field--free region below. While current sheets are a problem when they are used
to separate components in a magnetic atmosphere (see section \ref{forces}), in our model the 
current sheet is an integral part of the physics. It defines the boundary between the field free 
gaps and the magnetic field, and the forces associated with it serve to satisfy the 
condition of pressure balance at this interface. 

We emphasize that our primary goal is to investigate the consequences 
of field--free gaps below the surface on the magnetic field {\em above} the surface, not to provide 
a model for the {\tem shape of} gap itself, which is modeled rather crudely. In particular, it lacks the 
{\em cusp} (cf. Fig.\ \ref{gaps}) that will be present in a more realistic model.

\subsection{Properties of the model}
The figure shows how the field lines bend around the gap, and how this causes the field
lines to be further apart from each other above the gap. This predicts reduced field
strengths observed in this region. The figure does not show the component of the
field parallel to the filament, which is homogeneous and perpendicular to the plane of the
figure. 

The top of the gap would be the region we identify with the dark core of the filament. A horizontal 
field there, close to the continuum level, would also help to explain the Evershed flow. Because
of the rapid decline of density with height, this flow must be close to horizontal, covering
a vertical extent not exceeding a couple of pressure scale heights. This in turn {\em
requires that the optical depth unity surface of an observed filament does not deviate from
horizontal by more than about a pressure scale height}, at least in its outer penumbra where
the Evershed flow is observed. In our model, this is guaranteed because it identifies the
surface of the filament with the boundary between the penumbral magnetic field and the
convection zone below. Pressure balance across this interface determines its location in
depth. The magnetic pressure of the penumbra causes this depth to be depressed below the
normal photosphere by not more than about one scale height (somewhat more in the inner
penumbra where the field strength increases to umbral values). A possible model for the
force driving the Evershed flow is the `siphon effect' described  by Schmidt and Meyer.

The field becomes exactly horizontal just at the top of the gap, where the field lines 
following the sides of the gap meet. To be consistent with observations, this must happen
close to the $\tau=1$ layer.  This is also consistent with the heat flux requirement 
mentioned in the Introduction: The observed heat flux can be carried by radiation only if 
the surface of the gap is close to the continuum $\tau=1$ level (within a few tens of km). 
{\em This in turn implies that the field free gap must contribute to some extent to the 
formation of spectral lines}. This is an intriguing possibility for explaining the `stray light' 
contributions deduced from observations, as well as the very low field strengths found in the 
deepest layers in the inversions of Westendorp et al., Bellot Rubio et al. (2004), and Borrero 
et al. (2005) (cf. sect \ref{interp}).

In our analytic model, the shape of the gap is not realistic: its top is approximately circular
instead of a cusp as sketched in Fig.\ \ref{gaps}. This is because the analytic model
accounts for the pressure balance between the magnetic field and the gap only in an
approximate way. Accurate calculations of field configurations in pressure balance are
possible with the methods used for constructing sunspot equilibria (Jahn and Schmidt 1994),
but these are much more demanding.

We finally note that the magnetic field in the penumbra will of course not be exactly a potential 
field; on the one hand because it is not static, on the other because it must contain material to
provide the opacity seen in the lines. This gas exerts a pressure that varies with height, hence 
introducing some deviations from a potential field. Given the relatively small amounts of gas
needed to provide line opacity, these deviations can be small.

\begin{figure}
 \centering  \includegraphics[width=0.9\hsize]{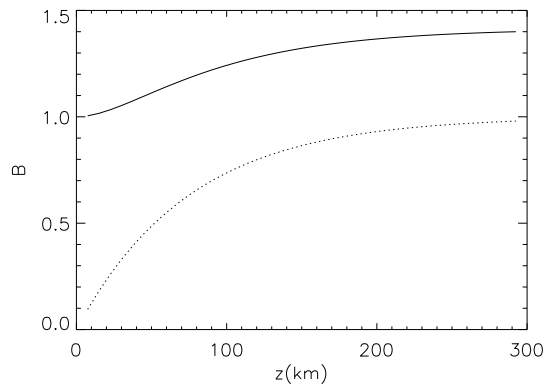}
 \caption{\small Variation of the vertical (dotted) and total field strength with height
 above the middle of the gaps shown in Fig.\ \ref{potfield}.}
\label{B-inc}
\end{figure}

Due to the gap, the magnetic field strength above it is reduced. This is evident from the 
shape of the field lines in Fig.\ \ref{potfield}, and shown explicitly in Fig.\ \ref{B-inc}.
This is also in agreement with the recent results of Langhans et al. (2005a).

\section{Discussion}

We have proposed that bright penumbral filaments are caused by field--free convection just below
the surface of the penumbra, and have discussed the theoretical and observational evidence for this
interpretation. The field free regions occur as radially oriented gaps in the field that makes up
the magnetic flux of the penumbra. We have argued that convection in these gaps between sheets of
magnetic field is the most plausible heating mechanism of the penumbra rather than convective flows
within a magnetic field.

The existence of such a magnetic field topology is supported by magnetogram data by Langhans et al.
(2005a), that consistently show polarization signal in the form of coherent structures extending
radially over nearly the entire penumbra.

The field--free gaps below the photosphere intrude into a magnetic fields that is nearly potential
(current--free) {\em above} the photosphere. This leads to a penumbral magnetic field with large
fluctuations on small scales both horizontally and vertically in inclination, azimuth angle,
and field strength. In particular, this potential field configuration always leads to nearly
horizontal magnetic fields around the tops of the field--free gaps. This explains the coexistence
of nearly horizontal magnetic fields with highly inclined fields between the bright filaments.

The model is a logical extension of the interpretation of umbral dots as caused by field--free gaps
just below the surface (Parker 1979a, Spruit 1981).  In this way it also provides a natural
explanation for the well-documented evolution of penumbral filaments into umbral dots.

Our model is distinct from the recent proposal by Thomas et al. {\tem (2002, see also Weiss et al. 
2004, Thomas and Weiss 2004, Tildesley and Weiss 2004) in which the penumbral structure is 
attributed to turbulent pumping by convection.  Observations show that spots are formed by 
a process of accumulation of previously erupted magnetic flux. Above the surface, this erupted flux is 
observed to spread out in the way expected from a magnetic field expanding into vacuum, i.e. it {\em 
overlies} the convection zone. The obvious problem how turbulent convection is to produce pumping 
in a magnetic field in which there is no convection was not addressed by Thomas et al. Instead, these 
authors supported their ideas by numerically simulating a case of a horizontal field already buried  
inside a convectively  unstable layer.

Thomas et al. propose that a field as seen in their simulations surrounds the spot.  As these 
simulations show, the pumping process would produce a dense mass of  `turbulent' magnetic field 
above the downward--pumped flux. Observationally, there is no evidence for such a field. While mixed 
polarities are seen as moving magnetic features (MMFs) in the moat flow around spots, the amount 
of (unsigned) flux present at any time in the form of MMFs is minute.

The source of the downward pumped magnetic field surrounding the spot, in the proposal of Thomas et 
al., are magnetic field lines of the penumbra. Dipping down of these fields lines into the convection zone 
at the edge of the penumbra  is proposed as the cause of the variations in field inclination seen in the 
penumbra. Such dipping down would be observed in the form opposite polarity flux surrounding the 
spot. This is ruled out by the fact that very little flux of polarity opposite to that of the spot is ever 
observed, either inside the penumbra or in the photosphere surrounding sunspots. 

Finally, even if an arbitrary amount of downward dipping flux were allowed at the edge, variations in 
field inclination produced by downward dipping at the edge of the penumbra could not possibly explain 
the observed inclination variations in the penumbra. Inclination variations as large as 45$^\circ$ 
persist all the way to the boundary with the umbra. An even greater difficulty is the very short vertical 
length scale of these variations, on the order of 200 km. Such a pattern of variations  cannot be produced 
by manipulating the field lines at the edge of the penumbra; they indicate a local origin.}

{\tem
In contrast, the model presented here is built on a mechanism that operates locally and
therefore has no problem explaining large inclination changes over a small height range anywhere 
in the penumbra.} In our model, these large fluctuations in inclination angle occur as a
natural consequence of field--free gaps below the surface, without {\tem the need for} any forces
acting on the gas above it. This removes the objections of Solanki et al. (1993) concerning the
large inclination changes inferred from Stokes spectra by Sanchez Almeida and Lites (1992). Our
model differs from the embedded flux tubes proposed by Solanki and Montavon (1993) to explain the
net circular polarization (NCP) measured in penumbrae.  We have shown in section \ref{forces} that
such tubes cannot survive in the observed layers.

Still, the embedded flux model has some similarity to ours by producing strong localized
perturbations in the magnetic field, within a few hundred km above the photosphere. Our model
differs mainly in the source of these perturbations, which we identify with field free regions
rather than embedded flux tubes. This similarity gives reason to believe that the model presented 
here will be able to reproduce the observed NCP, as was done successfully with the 
Solanki--Montavon model.

The model also has some conceptual similarities with that of Schlichenmaier et al.
(1998a,b). In both models, heat is carried to the observed surface from the convection zone below
by an elongated narrow structure, and some of the dynamics of the flows may also be similar in
both. Schlichenmaier's flux tube, however, is endowed with a magnetic field which we argue is
actually a hindrance rather than an asset, since it limits the heat flux it can carry.

Models like those proposed in the current literature have a natural pedigree in older thick
penumbra models (e.g. Danielson 1961) in which the temperature and magnetic fields are thought 
of as consisting of a smooth background with fluctuations on it. At a sufficiently rarified level,
such `magnetoconvection' type models can be interpreted as being related to a maximally
inhomogeneous model like our gappy penumbra. We hope to have shown with the present model, 
however, that such a interpretations do not lead very far, and in practice produce obstacles to physical 
understanding. {\tem A major obstacle invited by these interpretations is the intuitive mistake of viewing 
the low-$\beta$ region above the photosphere in the same terms as the high-$\beta$ regions below, 
namely as consisting of tubes that are dragged up and down by convective flows.}

The model proposed here naturally produces a horizontal magnetic field at some height above the
gap, thereby allowing for (but not requiring) the existence of horizontal Evershed flows in the
visible layers of the photosphere. Such flows are needed in combination with strong gradients or
discontinuities in the magnetic field in order to explain the NCP measured in penumbrae. In our
model, the heat supplied to these flows is provided by convection below the surface. This
also accounts for the fact that the intensity of penumbral filaments is rather uniform along their
length. 

While still resisting the temptation to claim an explanation of the Evershed flow, we note that our
model does put constraints on possible explanations. The field lines wrapping around the gap are
sufficiently horizontal to carry a plausible flow only over a finite distance. This implies that the
model can only be compatible with flows that are transient and rather local. A possibility that
suggests itself is that an Evershed flow may result from a {\em local} version of the
Schlichenmaier mechanism, operating in the field lines wrapping around the gap. Bundles of these
field lines could be heated by interaction with the hot gap. This could produce the same kind of
dynamics but on a much more local scale, namely in the boundary layer between the gap and the
surrounding field. That the Evershed flow, though smooth and steady on average, is in fact locally
transient is suggested by time series of Dopplergrams analyzed by Shine et al. (1994), Rimmele
(1994) and time series of spectra by Rouppe van der Voort (2003). These show the Evershed flow as
consisting of `velocity packets' repeating irregularly on a time scale on the order of 8--15
minutes.

Although our model has a horizontal magnetic field near the the top of the gap, 
this field is quite different from the long horizontal flux tubes simulated by Schlichenmaier et
al. (1998a,b). In our model, the origin of the horizontal magnetic field is field lines originating
{\em outside} the gap and wrapping over the gap towards the center of the filament, where the field
becomes aligned in the radial direction. This nearly horizontal field is associated with strong
variations in azimuth angle, as well as inclination angle. 

{\tem In the same way as}  the dark lanes in light bridges, explained by numerical simulations
of  Nordlund and Stein (2005), we identify the dark cores in penumbral filaments as surfaces of
enhanced gas pressure, occurring as the result of convection below the visible surface. Recent
magnetogram observations by Langhans et al. (2005b), indicate more horizontal magnetic fields in
the dark cores than in the lateral brightenings. They also report a significantly weaker
magnetogram signal in dark cores, which can be interpreted as weaker field strength. From
an analysis of spectra recorded in the Fe II 614.9 nm line, Bellot Rubio et al (2005) find
indications of slightly weaker magnetic fields in the dark cores as compared to the lateral
brightenings. This is consistent with our predictions (see Fig.\ \ref{B-inc}).

Recent work by Borrero et al. (2005) confirms the existence of a more horizontal field component
with lower than average strength, somewhat higher temperatures, and higher velocities. {\tem This 
is interpreted by these authors is in terms of the embedded flux tube model, and advanced as support
for it. The strength of this support is limited by the simplicity of the model used  to represent the flux 
tube (a horizontal flux sheet with neglect of perturbation in background field. The 
observational effects of these components are completely consistent also with our model.} 
This stresses again the ambiguities inherent in multi-component data inversions.

To resolve these ambiguities  more physics needs to be put into the models. In this context the 
Borrero et al. results give an interesting hint. The low field strength component in their model, 
envisaged by the authors as horizontally in equilibrium with the other component, must have a 
higher gas pressure. {\tem From their Figure 8, the difference in gas pressure is 
about $10^5$ erg/cm$^3$} in the middle of the penumbra. This implies that the low--field 
component is significantly {\em denser} than its environment (the slightly higher temperature has 
an opposite, but only marginal effect.) The two components can therefore not just float at the same 
level in the atmosphere. The weak--field component will {\em sink} unless supported from below by 
something that is able to exert a pressure of the order of  the photospheric value. In our model, 
this something is of course the field free gap.

We also note that inversions allowing for an unpolarized stray--light component usually come
up with an unexpectedly large fraction of unpolarized straylight in the penumbra (e.g Lites et al.
1993, Westendorp Plaza et al. 2001a) {\tem This is to be expected if the line is formed partly in
field-free regions. Hence it may well  be a signature of the field--free gaps  in 
our model. High stray--light fractions might  also be attributed to opposite polarities 
and consequent cancellation of Stokes profiles (Solanki 2003) if there were a reason for expecting 
such mixed polarities to be present}. 

As emphasized by e.g. Martinez Pillet (2000), the ambiguities 
associated with the inversion and interpretation of low--resolution spectropolarimetric data makes 
it very difficult to infer penumbral structure from the data alone. The more promising approach is to 
test sufficiently well--defined physical models such as our present model by forward modeling with 
detailed radiative transfer calculations.

What is clearly needed are spectropolarimetric data at high spatial resolution. Until such data are
available, inversions of low--resolution data incorporating a full "period" of our magnetic field
model, with the free parameters it allows, could already constitute a first feasibility
test. For future modeling, the strong 3D nature of the penumbra, shown clearly in SST images
recorded well away from disk center, should be accounted for. Such modelling will in particular
need to explain the strong differences in penumbra fine structure between the limb and the sun
center directions.

\acknowledgements{We thank {\AA}ke Nordlund, Bruce Lites, Kai Langhans, Mats L\"ofdahl, Luis Bellot 
Rubio, Jorge Sanchez and Sami Solanki for valuable comments on an earlier version of the text. HS 
thanks the Royal Swedish Academy of Sciences for support as visitor of the Telescope. The 
Swedish 1-m Solar Telescope is operated on the island of La Palma by the Royal Swedish Academy 
of Sciences in the Spanish Observatorio del Roque de los Muchachos of the Instituto de 
Astrof{\i}sica de Canarias. }

\end{document}